\documentclass{aa}  
\usepackage{natbib}
\usepackage{graphicx}
\usepackage{float}
\usepackage{txfonts}
\usepackage{aalongtable}
\begin{document}

\title{Pore evolution in interstellar ice analogues}
\subtitle{simulating the effects of temperature increase} 
\titlerunning{Pores in water ices.}
\author{S. Cazaux\inst{1}, J.-B. Bossa\inst{2}, H. Linnartz\inst{2}, A.G.G.M. Tielens\inst{3}} 
\offprints{cazaux@astro.rug.nl}
\institute{Kapteyn Astronomical Institute, University of Groningen, P.O. Box 800, NL 9700 AV Groningen, The Netherlands\\
\and
Sackler Laboratory for Astrophysics, Leiden Observatory, Leiden University, P.O. Box 9513, NL 2300 RA Leiden, The Netherlands\\
\and
Leiden Observatory, Leiden University, P.O. Box 9513, NL 2300 RA Leiden, The Netherlands}

\date{Received ; accepted }

\abstract {The level of porosity of interstellar ices - largely comprised of amorphous solid water (ASW) - contains clues on the trapping capacity of other volatile species and determines the surface accessibility that is needed for solid state reactions to take place.}{Our goal is to simulate the growth of amorphous water ice at low temperature (10~K) and to characterize the evolution of the porosity (and the specific surface area) as a function of temperature (from 10 to 120~K).} {Kinetic Monte Carlo simulations are used to mimic the formation and the thermal evolution of pores in amorphous water ice. We follow the accretion of gas-phase water molecules as well as their migration on surfaces with different grid sizes, both at the top growing layer and within the bulk.}{We show that the porosity characteristics change substantially in water ice as the temperature increases. The total surface of the pores decreases strongly while the total volume decreases only slightly for higher temperatures. This will decrease the overall reaction efficiency, but in parallel, small pores connect and merge, allowing trapped molecules to meet and react within the pores network, providing a pathway to increase the reaction efficiency. We introduce pore coalescence as a new solid state process that may boost the solid state formation of new molecules in space and has not been considered so far.}{}

\keywords{astrochemistry - ISM: abundances - ISM: molecules - stars: formation }

\maketitle

\section{Introduction}
Interstellar ices are formed in regions where UV photons are attenuated (\citealt{whittet2001}). In these regions, dust grains grow thick icy mantles (tens of monolayers) from accretion and deposition of atomic and molecular species onto their surfaces. This has been confirmed by observations of starless cores (\citealt{tafalla2006}) which show that the abundances of most gas-phase species suffer of a significant drop toward the core center. The missing gas species constitute the icy mantles that cover dust grains and that are largely composed of water (\citealt{williams1992}; \citealt{whittet1998}; \citealt{gibb2004}; \citealt{pontoppidan2004}; \citealt{boogert2008}). Other abundant species in the ices are CO$_2$, CO, CH$_3$OH and NH$_3$ (\citealt{gibb2004}). As a star forms and heats up the surrounding environment, ices in the vicinity undergo thermal processes and evaporate from the dust grains, hence delivering their constituents in the gas-phase. This star formation stage, called the hot core (massive stars) or hot corino (low mass stars) phase, exhibits a very complex chemistry rich in oxygen- and nitrogen-bearing molecules (\citealt{caselli1993}; \citealt{cazaux2003}), but also shows species such as formaldehyde (H$_{2}$CO) and methanol (CH$_{3}$OH). These species are precursors of more complex organic species which can be formed under thermal (\citealt{theule2013}) and Vacuum UV (VUV) processing (\citealt{oberg2009}) or upon continuing atom additions (\citealt{fedoseev2014}). Another property from these ices is that they can help dust grains to coagulate and then form bigger bodies, and therefore support the processes that will ultimately lead to planet formation (\citealt{ormel2007}).  

The majority of the studies investigating interstellar ices, both observationally and in the laboratory have been focusing on ice composition, rather than on ice structure (morphology). The ice composition reflects the chemical history, whereas the ice structure provides information on a number of physical processes: accretion, desorption, segregation and local radiation fields. The frost of water on interstellar grains in dense molecular clouds is thought to consist mainly of an amorphous water ice with trapped impurities (e.g., \citealt{tielens1987}). This is consistent with several theoretical studies where amorphous and porous ices are grown by a hit and stick mechanism (\citealt{cuppen2007}). In the laboratory, background deposition of gas-phase constituents on a cold substrate results in a multi-phase composite ice sample by taking the presence of pores into account, which are inevitably formed during growth (\citealt{brown1996}; \citealt{dohnalek2003}; \citealt{mate2012}; \citealt{bossa2014}). However, water ices formed experimentally from atomic hydrogen and oxygen present a compact structure (\citealt{oba2009}). Recent theoretical studies showed that the porosity of ices formed by surface reactions depends on the density of the gas (\citealt{garrod2013}). There is presently no observational evidence that interstellar ices present a porous structure because the OH dangling bonds of water on the surface of the pores, at 2.7 $\mu$m, has not been observed (\citealt{keane2001}; \citealt{gibb2004}). Moreover, many laboratory studies monitoring the 2.7 $\mu$m feature have shown that pores initially present in amorphous solid water (ASW) ices disappear with UV photon and ion irradiation (\citealt{palumbo2006}; \citealt{raut2007}; \citealt{palumbo2010}). The level of porosity is also found to decrease during an ASW ice growth (\citealt{mispelaer2013}). The authors explain the observed porosity evolution during an isothermal deposition by an ice relaxation process and/or by the increasing amount of water molecules forming the ice. Therefore, it seems that the lack of porosity in interstellar ices may be attributed to a change of ice morphology under external influences such as ion impact, VUV irradiation, H-atom bombardment, or during long time scale accretion.

In a number of recent experimental studies, however, it was proven that a lack of dangling OH bond does not necessarily exclude a remaining level of porosity in the ice (\citealt{raut2007}; \citealt{isokoski2014}) and special care needs to be taken to prove that water ice in space is fully compact. Porous ice -- even when largely compacted -- offers more area for surface dominated reactions, also within the bulk of the ice. Depending on the structure of the pores -- many small versus a few larger pores, connected or not -- this will affect the catalytic potential of an icy dust grain. Recent studies (\citealt{bossa2012},  \citealt{isokoski2014}) have focused on measuring the compaction of an ice upon heating. This allows to conclude on the level of porosity in an ice, but not on the actual pore size. 

Experiments performed by \cite{bossa2012} and \cite{isokoski2014} have shown that the thickness of a background deposited porous ASW sample decreases by about 12 $\pm$1$\%$ (upper limit) upon heating from 20 to 120 K. The thickness decrease of less-porous ASW is smaller, and negligible for crystalline solid water. While this compaction was first attributed to a loss of porosity, \cite{isokoski2014} found that pores do survive upon warming up and that the porosity of a porous ice initially deposited at 20~K and then annealed to 120~K is still around 20$\%$. The porosity of water ice is related to the temperature at which ices are deposited (but also to VUV and ion irradiation). The goal of the present study is to simulate the temperature evolution of porosity in an interstellar ice using a theoretical approach.

ASW exhibits tetrahedral H-bonding structure. This structure was first modeled by \cite{polk1973} with amorphous Si or Ge using a concept based on a tetrahedrally coordinated random-network structure. Each atom has a first coordination number of four, and only a small variation in the nearest neighbour distance is allowed. Non-crystallinity is due to variations in the tetrahedral bond angle and rotations along bonds. These models were extended in order to simulate the scattering of X-rays and neutrons on amorphous water ice  (\citealt{boutron1975}). 

The mobility of water molecules in amorphous water ices has been theoretically treated by \cite{zhdanov2000} using a ballistic deposition (the water molecules hit and can form at least one bond) followed by a defined number of jumps to near neighbour sites, the direction of the jumps being chosen randomly. These simulations, using a Monte Carlo approach, consider diffusion, crystallization and desorption of water in porous amorphous films in order to quantify the desorption processes. In the liquid phase, \cite{laage2006} have used molecular dynamic calculations and have shown that water molecules break and reform hydrogen bonds, and that this process is accompanied by a reorientation mechanism which involves large-amplitude angular jumps, rather than a sequence of small diffusive steps. The diffusion in the solid-phase through breaking and reforming bonds depends on the hydrogen bond strengths of the molecules in the ices. In a recent study \cite{hus2012}, using density functional theory, have shown that hydrogen bond strengths increase or decrease linearly with the number of water molecules in their surrounding. A simple empirical linear relationship was discovered between maximum hydrogen bond strength and the number of water molecules in the local environment. The hydrogen bond strengths in the study by \cite{hus2012} range from 4.96 kcal/mol (2490~K) to 9.41 kcal/mol (4735~K), depending on the number of donors and acceptors in the local environment. 


In this study, we use Monte Carlo simulations to follow the migration of water molecules within the ices, and the thermal evolution of pores with temperature. Our simulations show that when ices become warmer, the water molecules rearrange and the empty spaces between the agglomerate of strongly bound water molecules expand. This implies that pores grow and merge as the temperature of the ices increases. While the merging/growth of pores in ice was previously suspected (\citealt{raut2007}; \citealt{isokoski2014}), this work highlights for the first time this behaviour in a quantitative manner.

\section{Modeling amorphous solid water}
We use a step-by-step Monte Carlo simulation to follow the formation of amorphous water ices as well as the temperature evolution of the pores within the ices. Water molecules originating from the gas-phase arrive at a random time and location and follow a random walk within the ice. The arrival time depends on the rate at which gas species collide with the grain. The time for accretion and diffusion of the water molecules is defined in the next sections. 

\subsection{Structure of the ices: assumptions}
\begin{figure*}[!ht]
\includegraphics[scale=0.35]{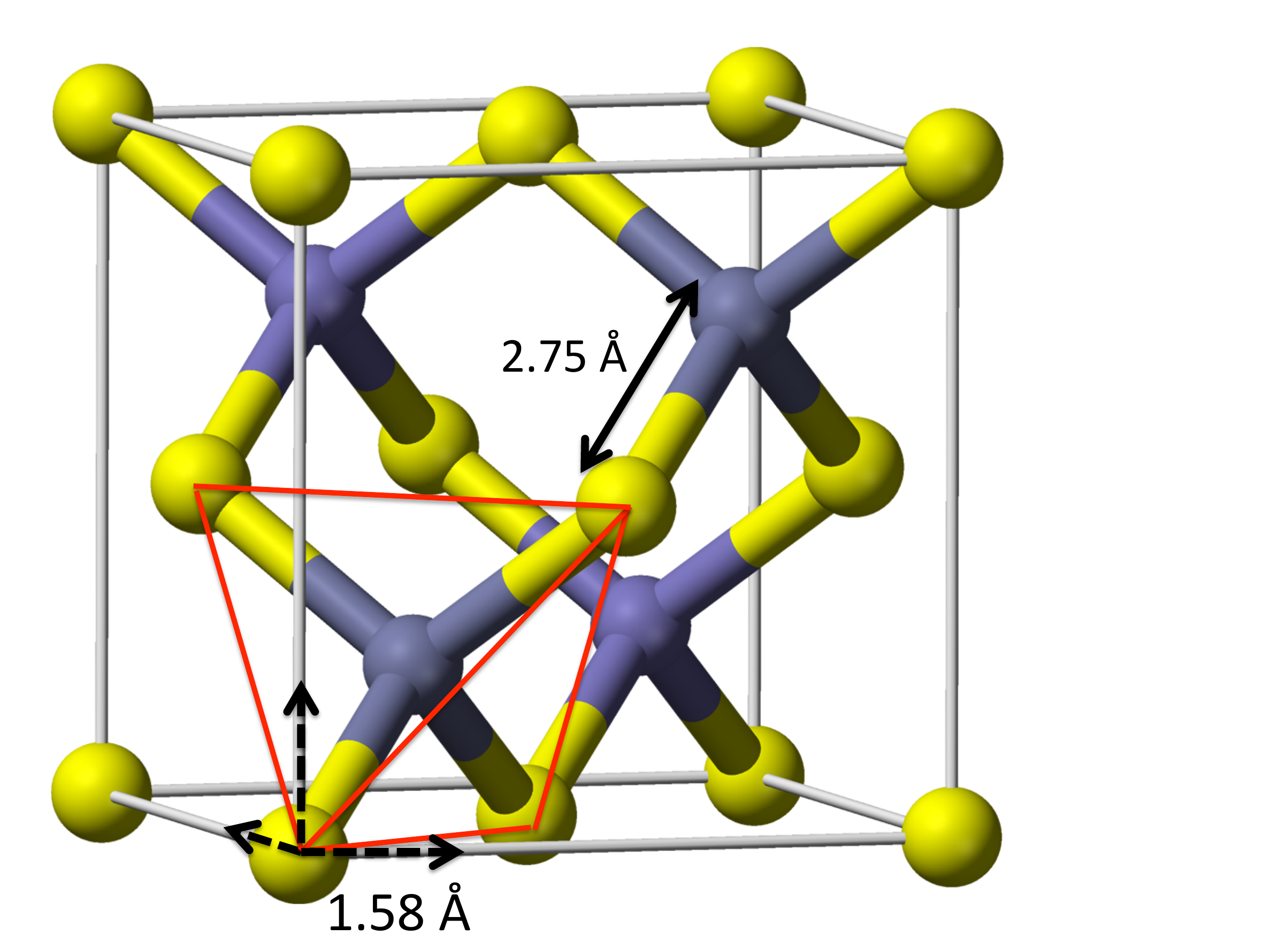}
\includegraphics[scale=0.25]{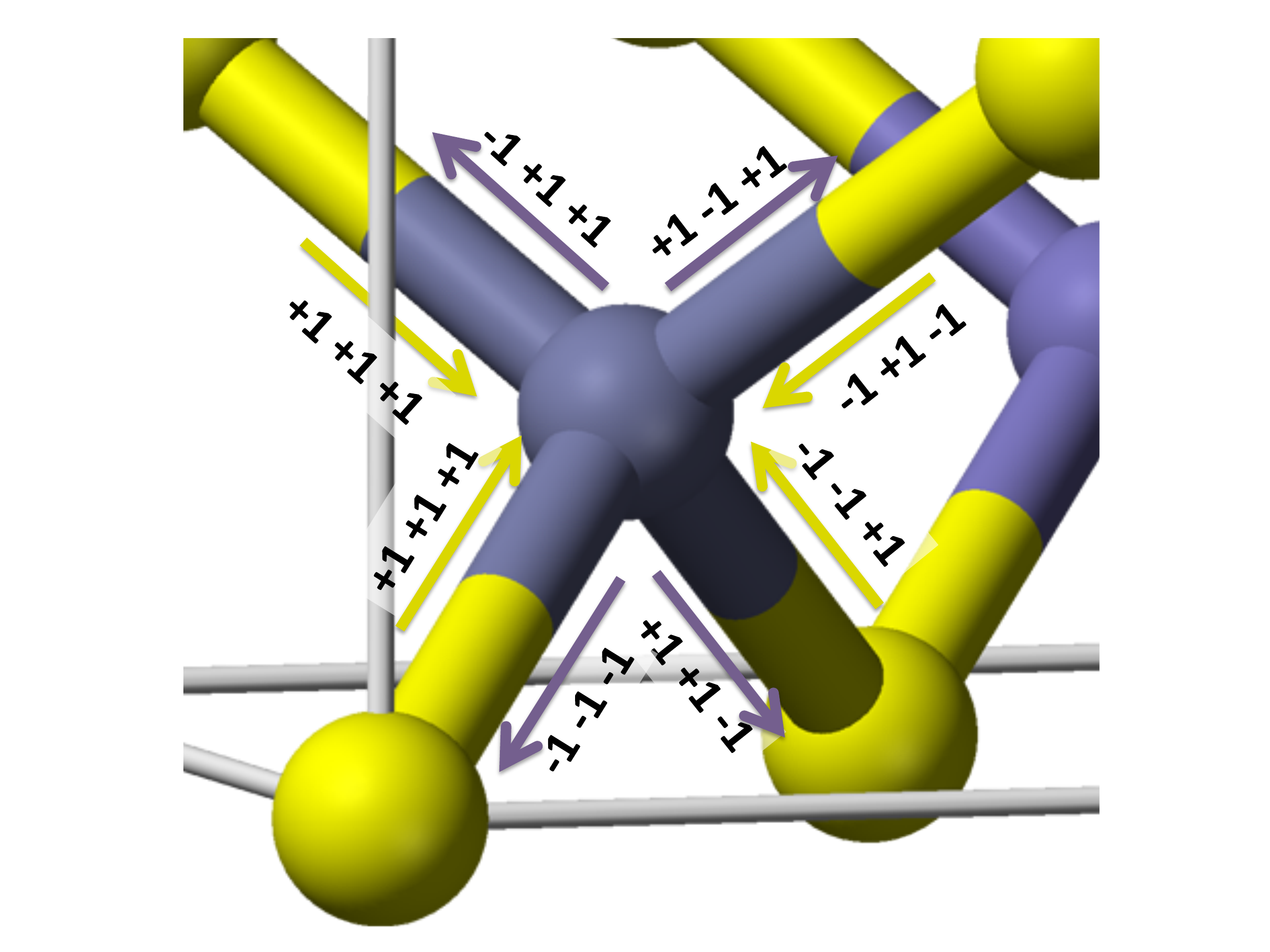}
\caption{Left panel: structure of water ices in a 4$\times$4$\times$4 grid. Each knot represent the position of an oxygen atom with odd vertical coordinates (yellow) and with even vertical coordinates (blue). This box illustrates the possible positions of water molecules in our grid model. The grey lines show the tetrahedrons in our grid that are used to calculate surface and volume.  Right panel: Water molecules with odd vertical coordinates (yellow) can move in the grid following the yellow arrows, while water molecules with even coordinates (blue) can move in the grid following the blue arrows. }
\label{grid}
\end{figure*}

Low-density amorphous ice consists mainly of the four coordinated tetrahedrally ordered water molecules (\citealt{brovchenko2006} and references therein). In order to model amorphous water ice, we define our grid model to allow the water molecules to be organised as tetrahedrons. We therefore pre-define which sites of the grid can be occupied and which sites should stay vacant. We concentrate on the molecular centers, determined by the oxygen positions, and consider a four - coordinated structure since it is amorphous. An example of our grid is shown in figure~\ref{grid}, where each knot represents an oxygen atom, and the different colors illustrate odd and even grid cell heights. The bonds between the oxygens represent the hydrogen bonds. This figure shows how water molecules are organised in a 4$\times$4$\times$4 grid, which means that only positions at the knots of the grid can be occupied. We consider the distance between two oxygens to be 2.75 \AA\ (\citealt{boutron1975}) which defines our grid cells with a size of 1.58 \AA (which corresponds to the diagonal of the surface of a grid cell d=2.75$\times \sin(\frac{109}{2})$=2.24 \AA, divided by $\sqrt{2}$). The surface and volume of the cell is described as tetrahedrons (grey lines in figure~\ref{grid}, left panel). Each tetrahedron of our grid has a edge length of 4.5 \AA (2$\times$d), a surface of 4$\times$ 8.7 \AA$^2$ (each face has a surface of $\frac{\sqrt(3)}{4}\times d^2$ and a volume of 10.75 \AA$^3$ ($\frac{1}{6\sqrt(2)}\times d^3$). In our simulations, we consider three different grid sizes, with a base of 40$\times$40, 60$\times$60 and 100$\times$100 grid cells. We set the height of the grid at 150, so that we can study ices of different thicknesses.

\subsection{Accretion}
Water molecules from the gas-phase arrive on the grid, and can be bound to water molecules already present in the grid through hydrogen bonding. The accretion rate (in s$^{-1}$), depends on the density of the species, their velocity and the cross section of  the dust surface, and can be written as:
\begin{equation}
R_{acc} = n_{\rm{H_2O}} v_{\rm{H_2O}}  \sigma  S, 
\end{equation}
where $v_{\rm{H_2O}} \sim 0.43 \sqrt{\frac{T_{gas}}{100}}$ \rm{km~s$^{-1}$ is the thermal velocity, S the sticking coefficient that we consider to be unity for the low temperature taken as a starting point in this study. We also assume that every water molecule arriving on the substrate (grid not covered) or on another water molecule has a probability of 100$\%$ to stick. $\sigma$ is the cross section of the surface and directly scales with the size of the grid we use for the simulations as 1.58$\times\ n_{sites}^2$ \AA$^2$. n$_{\rm{H_2O}}$ is the density of water molecules in the gas in cm$^{-3}$, we here use 10$^{10}$ cm$^{-3}$ in our computations to mimic experimental conditions. In the experiments, ice samples are grown by background deposition in the high-vacuum chamber with a rate of 0.4 nm s$^{-1}$. Note that for the deposition temperature considered in this study, the water molecules stick to each other and do not diffuse and re-organise. This implies that the density of water molecules in the gas does not affect the results of our simulations.)}\rm\ Water molecules arrive in a random coordinate onto the grid, but can occupy only pre-determined positions, as described in the previous subsection. As the water molecules arrive on a location of the grid,  hydrogen bonds can be made with already bound water molecules. In this sense, we do not take into account chemical processes and we assume that the ice matrix is constructed upon accretion, rather than through oxygen allotrope hydrogenation (\citealt{dulieu2010}; \citealt{romanzin2011}; \citealt{oba2009}; \citealt{ioppolo2008}).

\subsection{Mobility of water molecules within the ice}
The binding energy of a water molecule increases with its number of neighbours and is estimated at 0.22 eV per hydrogen bond (\citealt{brill1967}; \citealt{isaacs1999}; \citealt{dartois2013}). In fact, the hydrogen bond strength depends on the number of water molecules involved as donors and acceptors in the neighborhood (\citealt{hus2012}). In this study we assume that the binding energies increase linearly with the number of neighbours, as also previously assumed by \cite{cuppen2007}, so that water molecules surrounded by one neighbour have a binding energy of 0.22 eV and water molecules surrounded by four neighbours have a binding energy of 0.88 eV.  The water molecules present in the grid can move from one site to another one as shown in figure~\ref{grid}, right panel. The water molecules located in odd vertical coordinates (yellow knots) can change their coordinates as (+1, -1, -1); (+1, +1, +1); (-1, +1, -1) and (-1, -1, +1), represented with blue arrows, while the molecules located at even vertical coordinates (blue knots) can change in the opposite direction (red arrows).  According to their positions, the molecules can move in 4 different directions in unoccupied sites. The diffusion rate, in s$^{-1}$, for a water molecule to move from one site to another can be written as:
\begin{equation}
\alpha({n}) = \nu \exp\left[{-0.4\times\frac{nn\ E_b}{T}}\right], 
\end{equation}
where $\nu$ is the vibrational frequency of a water molecule in its site (that we consider as 10$^{12}$ s$^{-1}$), E$_b$ is the energy of a single hydrogen bond, and \it{nn}\rm\ the number of neighbours. Here we consider the activation energy for diffusion to be 40$\%$ of the binding energy. This value is comparable to the value of 30$\%$ derived from experiments of water-on-water diffusion (\citealt{collings2003}) and 50$\%$ used in \cite{cuppen2007}. This number is however very uncertain and can reach 90$\%$ (\citealt{barzel2007}).

\subsection{Evaporation}
The species present at the top layers of the ices can return into the gas-phase because they thermally evaporate. This evaporation rate depends on the binding energy of the water molecules and is therefore directly dependent on the number of neighbours \it{nn}\rm. The evaporation rate (in  s$^{-1}$) of one species with \it nn\rm\ neighbours can therefore be written: 
\begin{equation}
 evap= \nu \exp({-\frac{nn\ E_b}{T}}). 
\end{equation}
We do allow evaporation of water molecules \it in\rm\ the ice, and consider that these molecules can be re-adsorbed at another location of the ice or be released in the gas-phase. The molecules desorbing from the ice can travel through the pores to meet another adsorbed water molecule (on the other side of the pore) and recreate a H-bound. Note that water molecules at the top layer of the ice have less neighbours and therefore have a higher probability to thermally evaporate than water molecules in the bulk (equation 3 with \it nn\rm\ smaller for molecules on top layers).

\section{Simulations}
In this section we describe several simulations to study the evolution of water ices with increasing temperature. Our calculations consider different grid sizes with different thicknesses. We follow the position of each molecule as well as the number of H-bonds with it neighbours. Therefore, molecules can have one to four neighbours and their total binding energies increase accordingly. Since we are interested in the porosity of the ices, we quantify the size of the pores as well as the total cross section. To do so, we tag every empty grid space with a number that corresponds to the number of grid cells around this cell which are empty. In our simulations, these tags can range from 0 (an empty spot with at least one water molecule as neighbour) to 6 (an empty spot with no water molecule present at a distance of six grid cells around, which means that the hole is the center of a sphere with six grid cells radius). This provides a direct measurement of the total pore volume, the total surface area but also the individual pore volume.

\subsection{Ice and pores thermal evolution}
\begin{figure*}[!ht]
\includegraphics[scale=0.35]{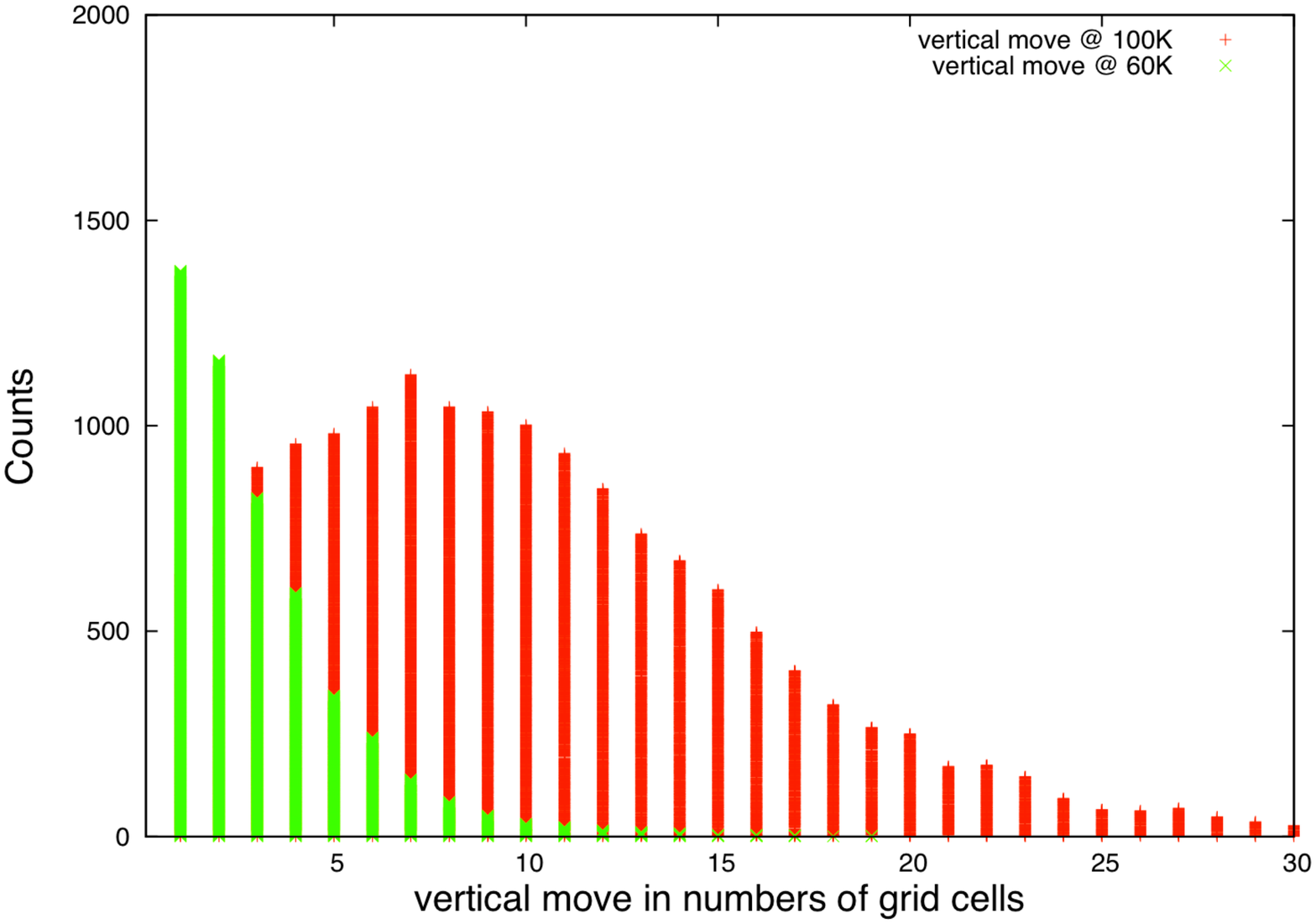}
\caption{Mobility of water molecules in the grid in vertical direction during deposition and warming up until 60~K (green) and warming up until 100~K (red). The number of counts is the number of molecules travelling x grid cells vertically. }
\label{move}
\end{figure*}

\begin{figure*}[!ht]
\includegraphics[width=0.33\textwidth]{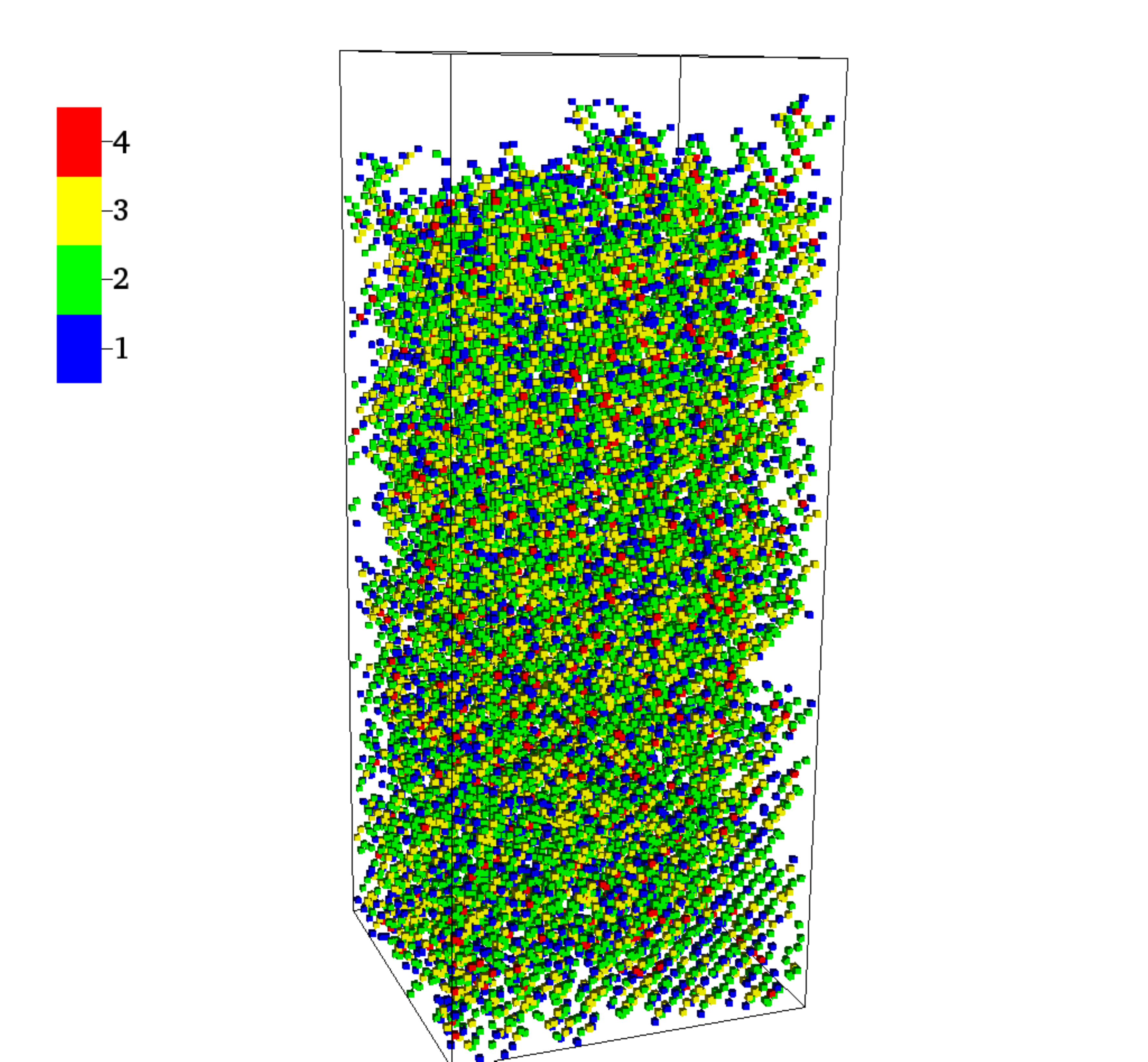}
\includegraphics[width=0.33\textwidth]{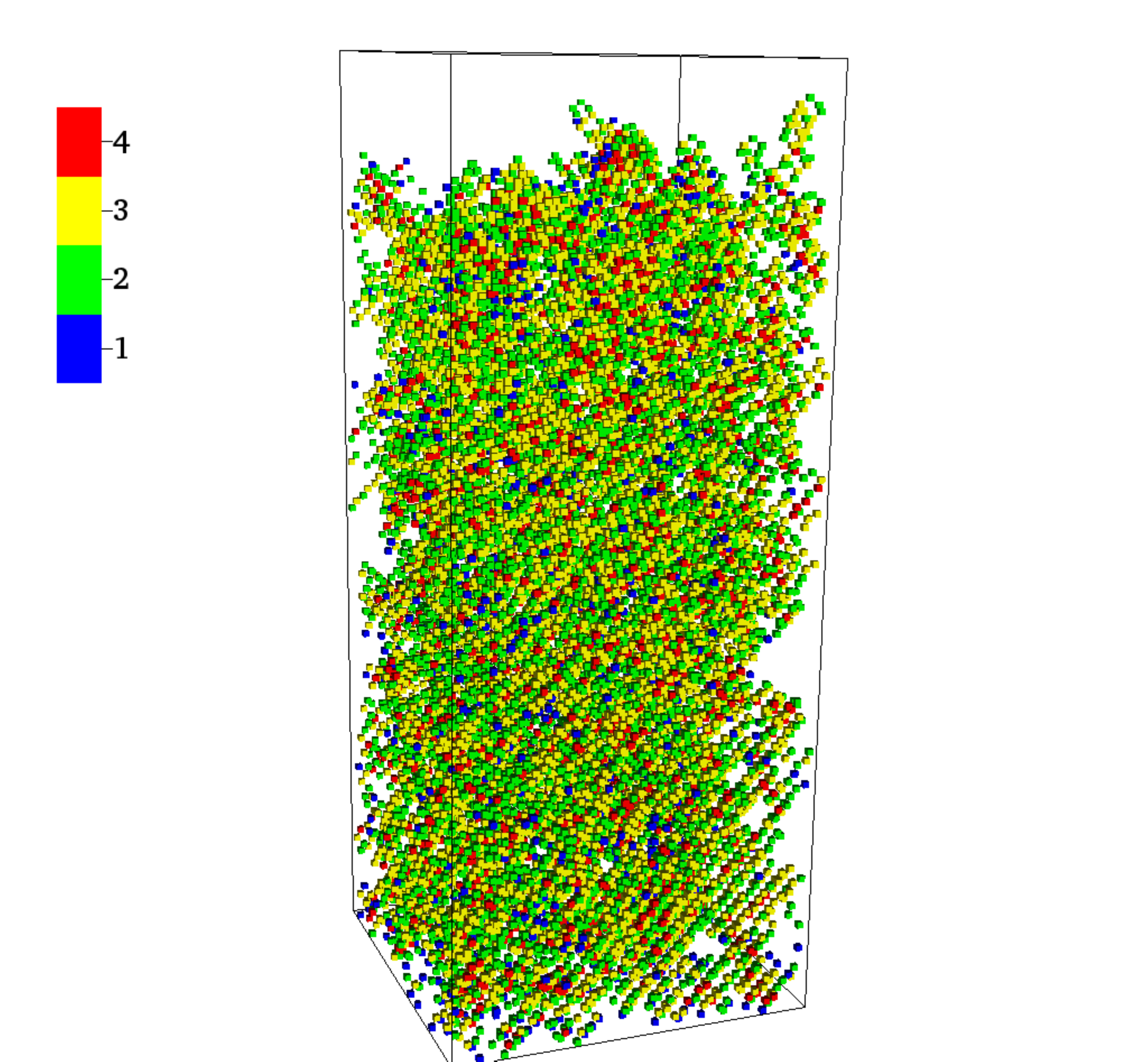}
\includegraphics[width=0.33\textwidth]{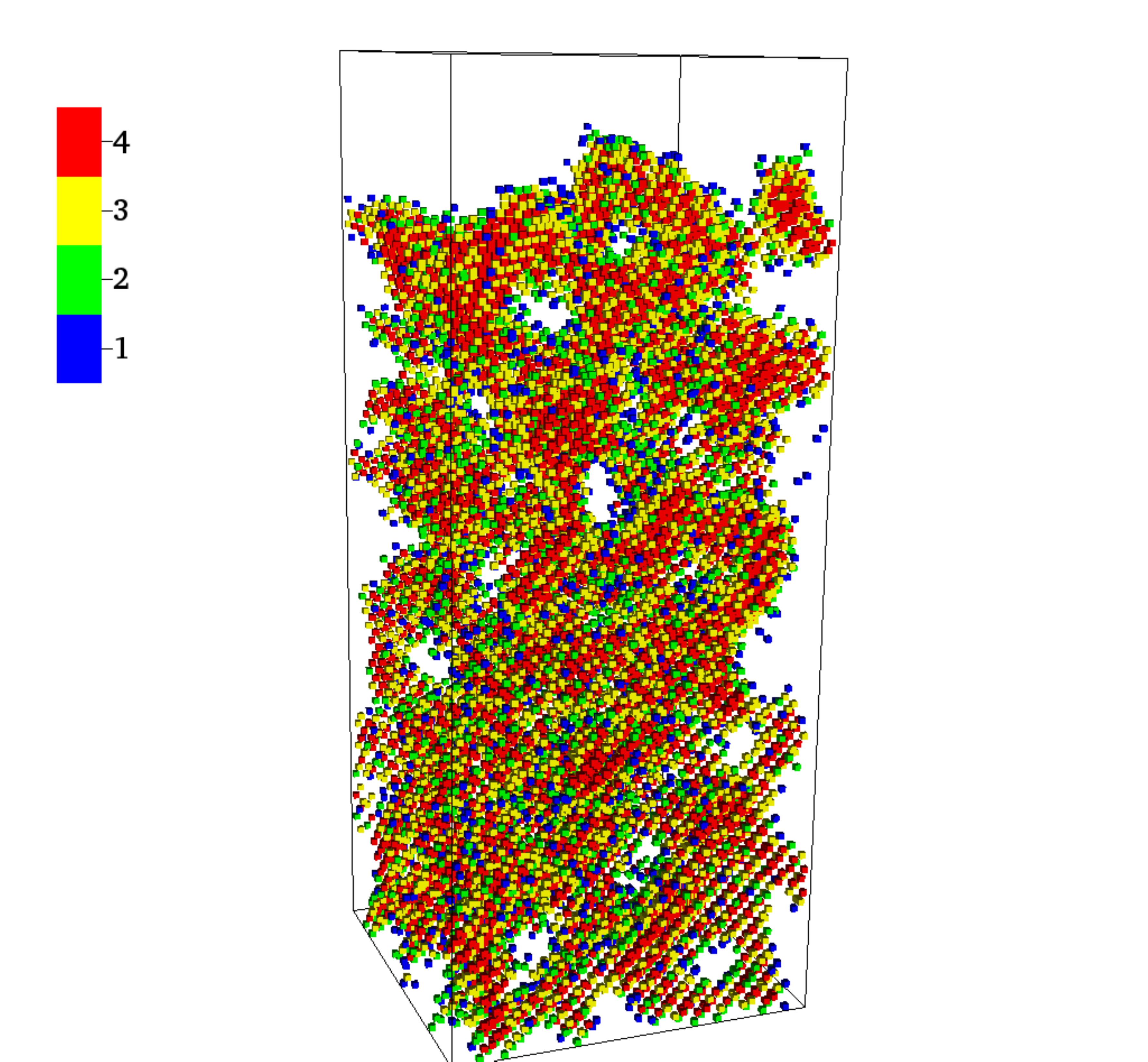}
\caption{Structure of the ice in a 60$\times$60$\times$150 grid at 10~K (left), 60~K (middle) and 100~K (right). The temperature is initially set at 10K, and is increased at a rate of 2K/minute to mimic laboratory conditions. The colors represent the number of neighbours of each molecule.}
\label{grid1}
\end{figure*}

\begin{figure*}[h]
\includegraphics[width=0.33\textwidth]{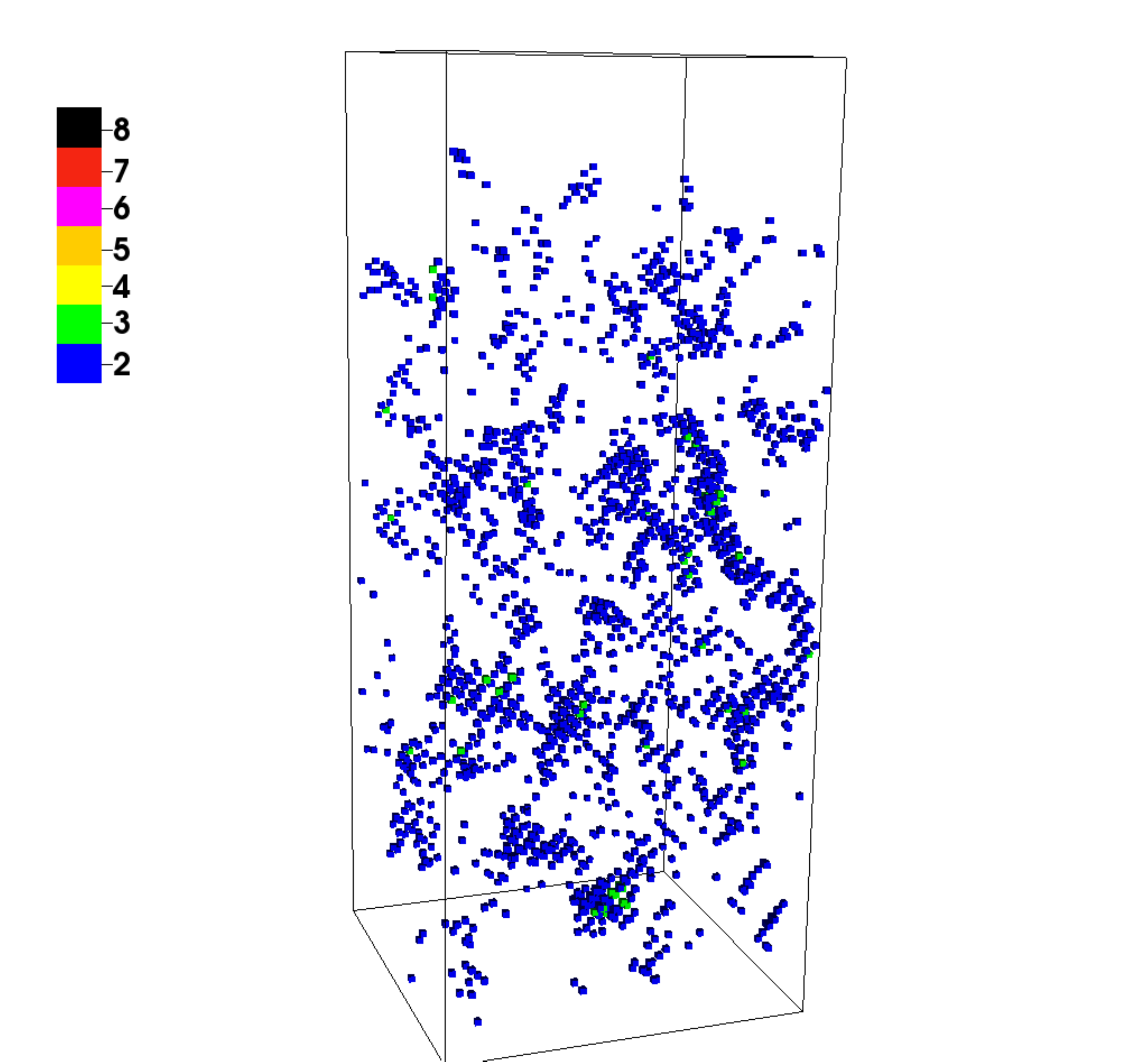}
\includegraphics[width=0.33\textwidth]{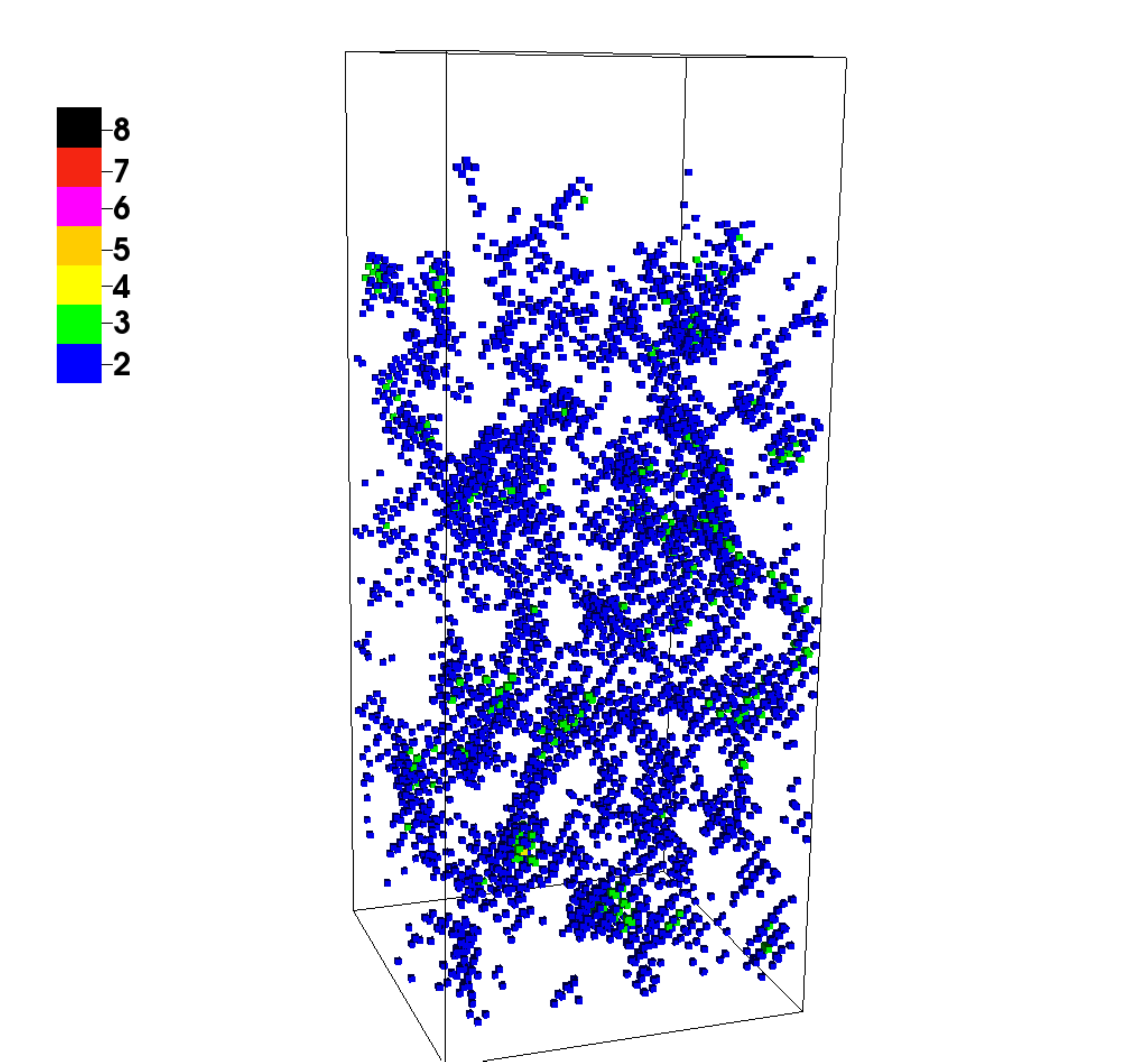}
\includegraphics[width=0.33\textwidth]{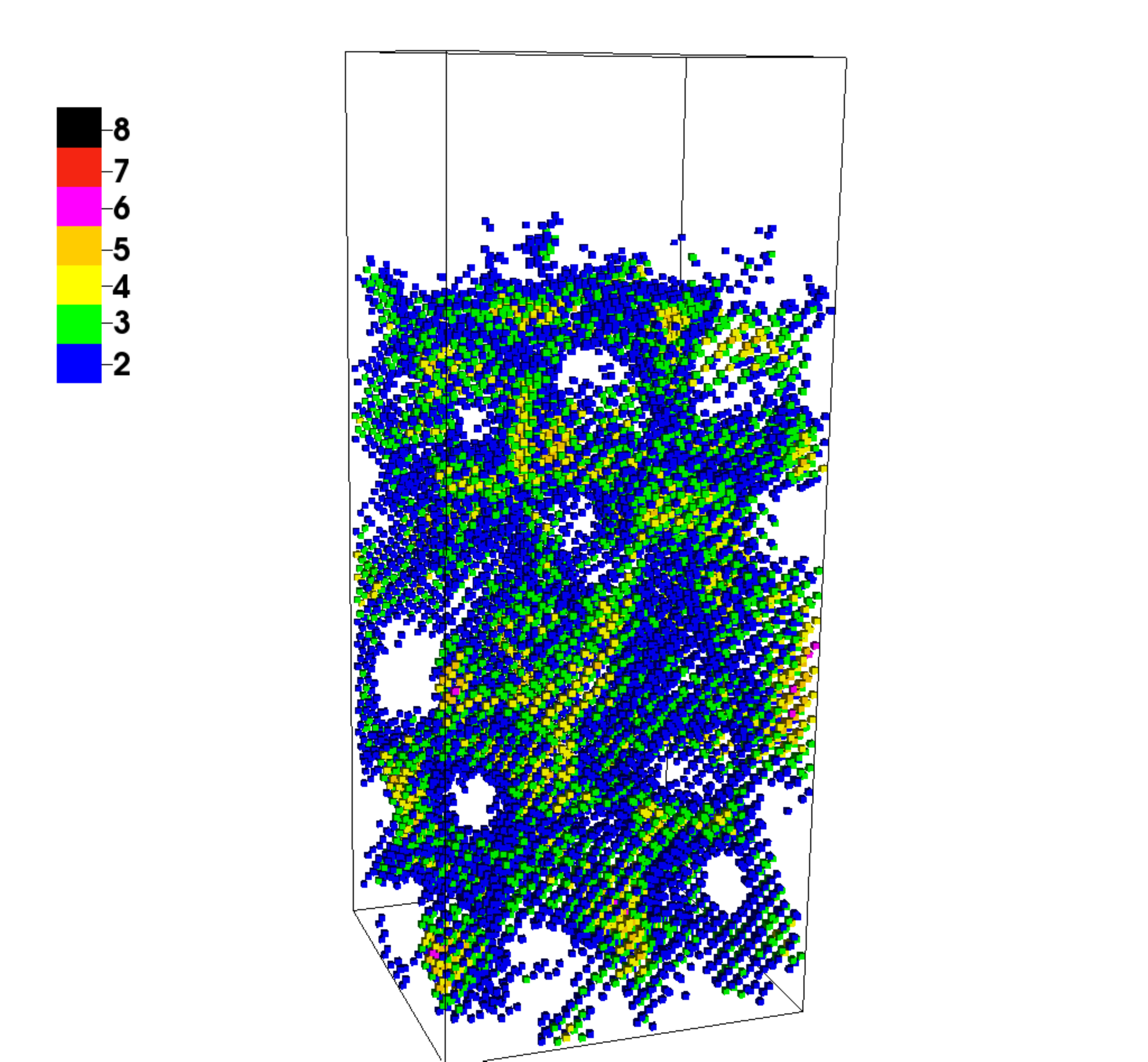}
\caption{Distribution of the pores in a 60$\times$60$\times$150 grid at 10~K (left), 60~K (middle) and 100~K (right). The temperature is initially set at 10K, and is increased at a rate of 2K/minute to mimic laboratory conditions. Each dot shows an empty space with a tag n (color scale) corresponding to the number of empty cells in a radius n around the cell n. This figure is the negative image of figure~\ref{grid1}. }
\label{grid2}
\end{figure*}

We have performed several simulations on different grids with sizes of 40$\times$40, 60$\times$60 and 100$\times$100. The first motivation of this work is to reproduce the thickness decrease of porous ASW upon thermal annealing. We therefore consider temperatures during deposition (10~K) and heating rates (2~K/min) identical to the values used in the laboratory studies. The water ices are deposited at 10~K and are subsequently warmed up at a rate of 2~K per minute until a temperature of 120~K is reached. 
Since we assume that the binding energy of each water molecule is proportional to the number of H-bonded neighbours, water molecules with only one neighbour are the most mobile and can break and reform bonds to arrive in more stable configurations (more neighbours). As the temperature increases, molecules with two and then three neighbours also become mobile. This re-arrangement of molecules is monitored as a function of temperature. In figure ~\ref{move}, we show the number of vertical grid cells, water molecules visit during warming of the ice. Molecules scout a few grid cells (two to five) at temperatures up to 60~K, whereas they can travel through much higher numbers (6-20) at 100~K. Therefore, as the temperature increases, the water molecules travel much larger distances, which allows them to rearrange in a more stable configuration. In figure~\ref{grid1} we show the 3D arrangement of water molecules in the grid of a size 60$\times$60. The left panel presents the ice at 10~K. The water molecules are mostly bound with one or two neighbours, and these molecules fill up the grid homogeneously. As the temperature increases, the molecules re-arrange in order to be in a more stable configuration. Most of the molecules have two to three neighbours at 60~K (middle panel). At 100~K, as shown in the right panel, the water molecules in the ice are very strongly bounded and have in majority three to four neighbours. The ice is not as homogeneous as at low temperatures, and large empty spaces do appear which represent large pores. The pores and their evolution as function of temperature are presented in figure~\ref{grid2}.  This figure is a negative of figure~\ref{grid1}. This figure shows the distribution of pores during the increase of temperature, and illustrates their growth from 10~K (left panel) to 60~K (middle panel) and to 100~K (right panel). In these figures, we represent the pores with a tag of 2 (no water molecules are present at a radius of two grid cells around) until 8 (no water molecules at a radius of height grid cells around). The color shows the tags associated with each empty cell. At low temperatures, the ice is filled with holes and water molecules. As the temperature increases, the water molecules get re-organised to be in a more stable configuration, which leads to the creation of larger empty spaces, the pores. These pores become larger and can reach radii of a few \AA\ at 60~K. At even larger temperatures, the pores grow and connect to each other. The sizes of the pores can reach radii of $\sim$15 \AA. Note that these radii correspond to empty spheres present in the ices (which is how we defined pore sizes), while pores can be extended to empty volumes as represented in figure~\ref{grid2}. Therefore, our values represent lower limits for the volume of the pores. However, the sizes we computed for individual  pores are in good agreement with the diameter $\le$20 \AA\ found in the literature (\citealt{raut2007}).

\begin{figure*}[h]
\includegraphics[width=0.34\textwidth,angle=-90]{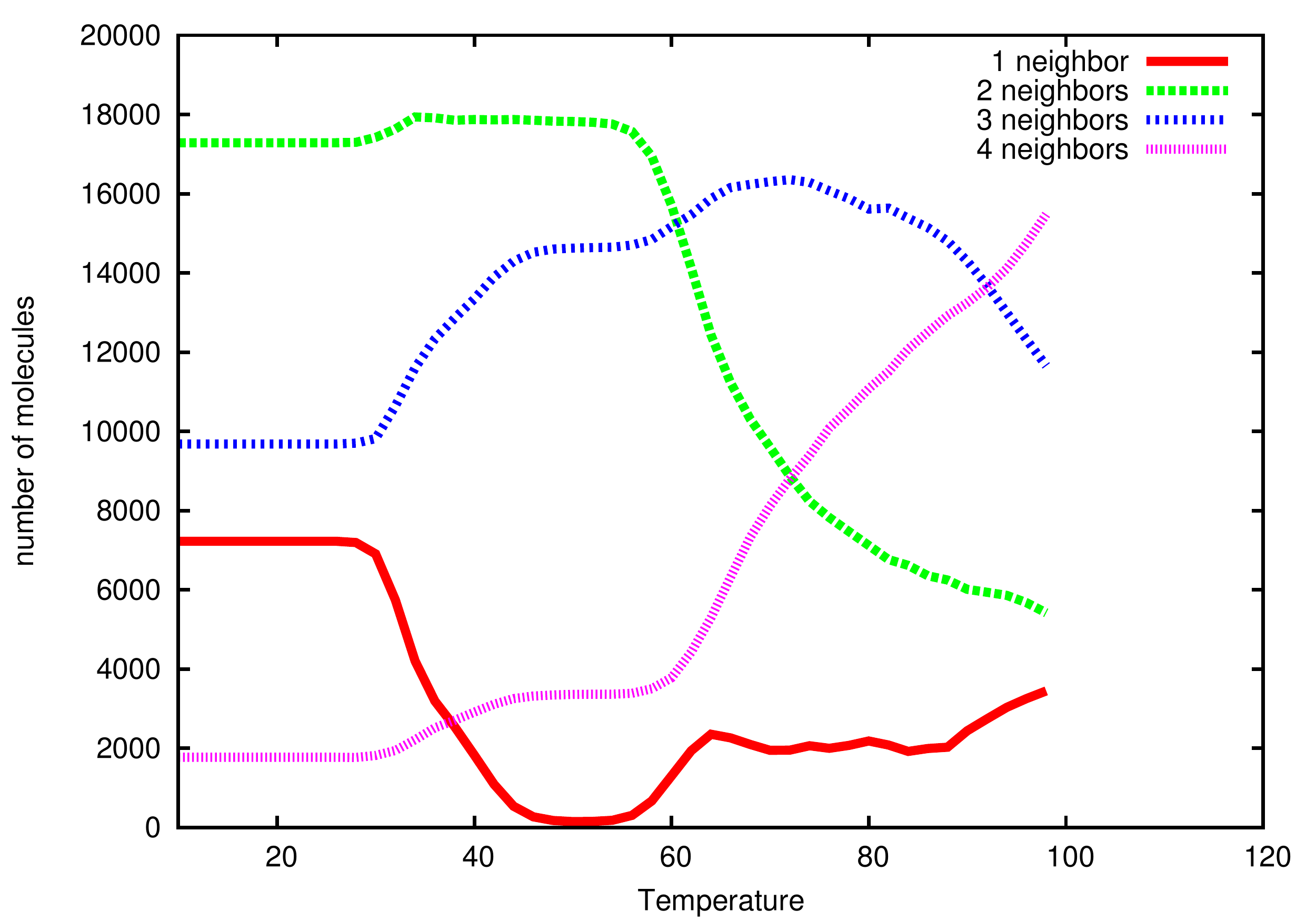}
\includegraphics[width=0.34\textwidth,angle=-90]{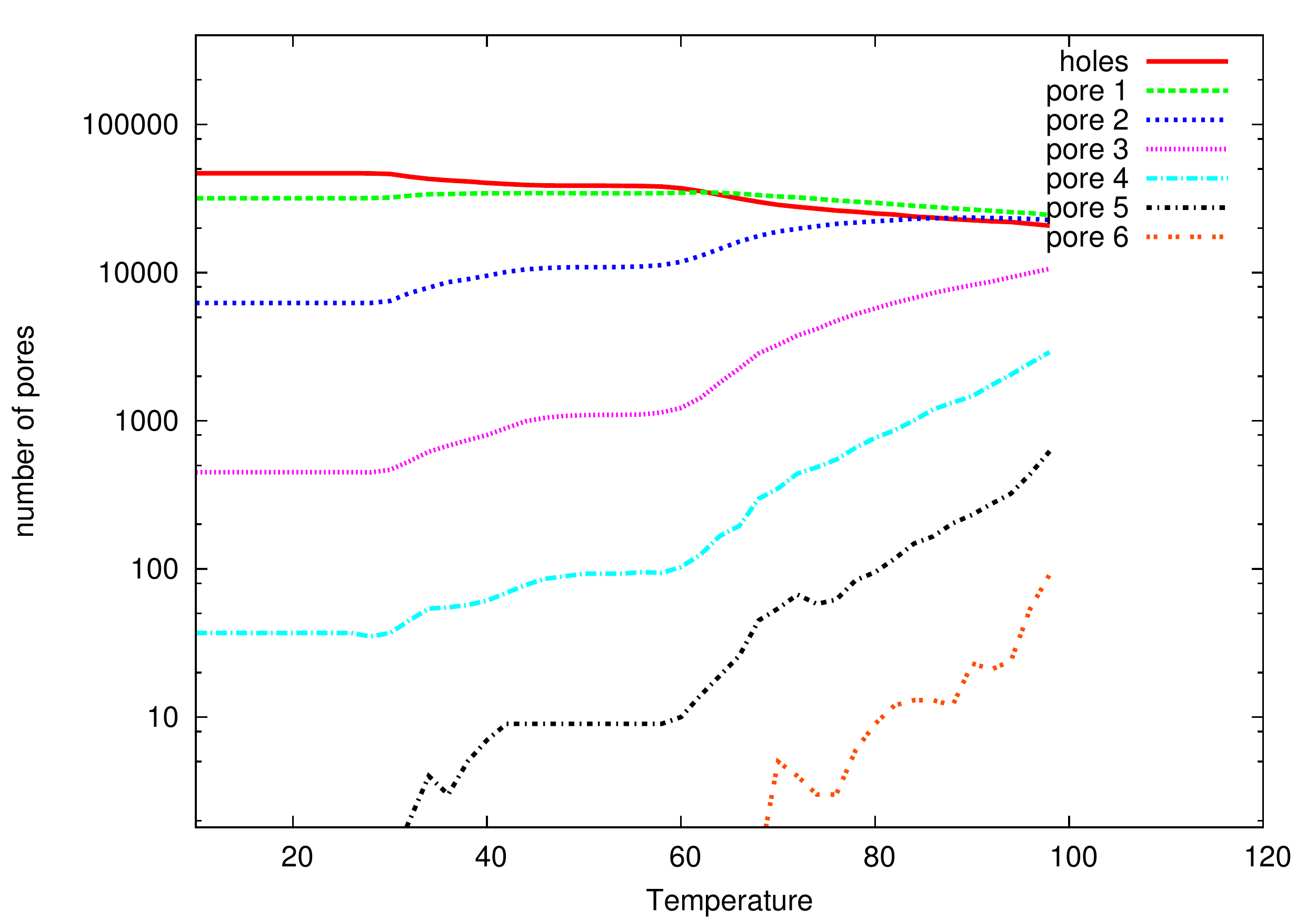}
\caption{Left panel: number of water molecules with a certain number of neighbours in a 60$\times$60 grid as a function of temperature. The temperature is initially set at 10K, and is increased at a rate of 2K/minute to mimic laboratory conditions. The total number of water molecules is 18000. Right panel: number of pores as a function of temperature. The numbers n in "pore n" indicate the radius of empty spheres in number of grid cells. These numbers are directly related to the size of the pores (pore 6 has a radius of 6$\times$ 2.75 \AA). }
\label{nnpores}
\end{figure*}

In figure~\ref{nnpores}, left panel, this finding is summarized quantitatively and we report the evolution of the number of neighbours of the water molecules as function of the temperature. As mentioned above, at low temperatures water molecules have in majority two neighbours. As the temperature increases, all molecules with two neighbours (or less) will diffuse in the ice until they have at least three bonds. At temperatures higher than 90~K, the molecules have in majority four neighbours and are very strongly bond to each other. The corresponding evolution of the pores in the ices is reported in figure ~\ref{nnpores}, right panel. While the ice is dominated by little holes (pores with radii of 1 or less) at low temperatures ($\le$ 4 \AA\ radii), the re-organisation of the ice implies that these holes become connected to form bigger entities. The size of the pores ( radii $\ge$2) grows with temperature, and the ice becomes dominated by larger empty spaces.

\subsection{Pore volume and surface area}
\begin{figure*}[h]
\includegraphics[width=0.34\textwidth,angle=-90]{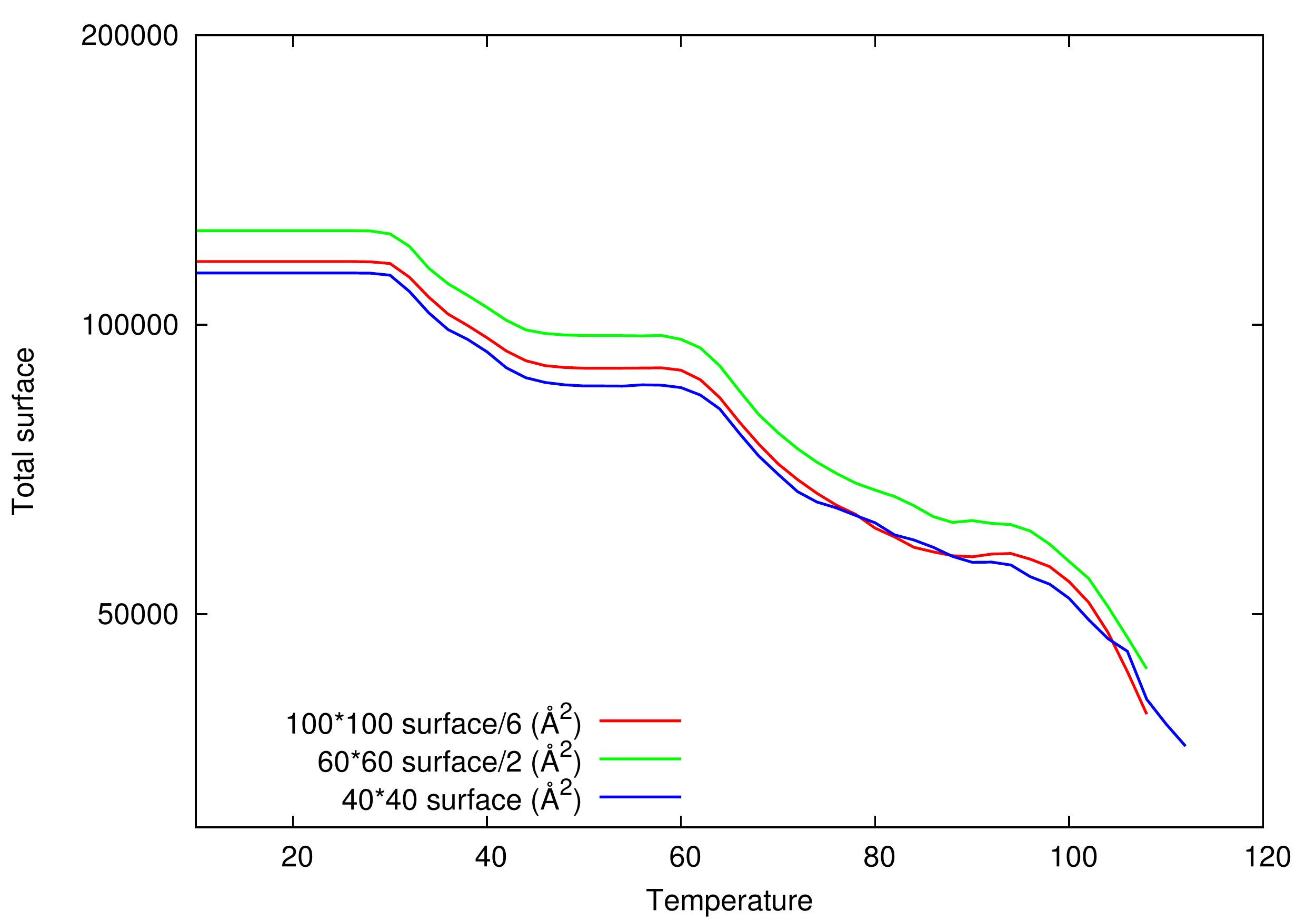}
\includegraphics[width=0.34\textwidth,angle=-90]{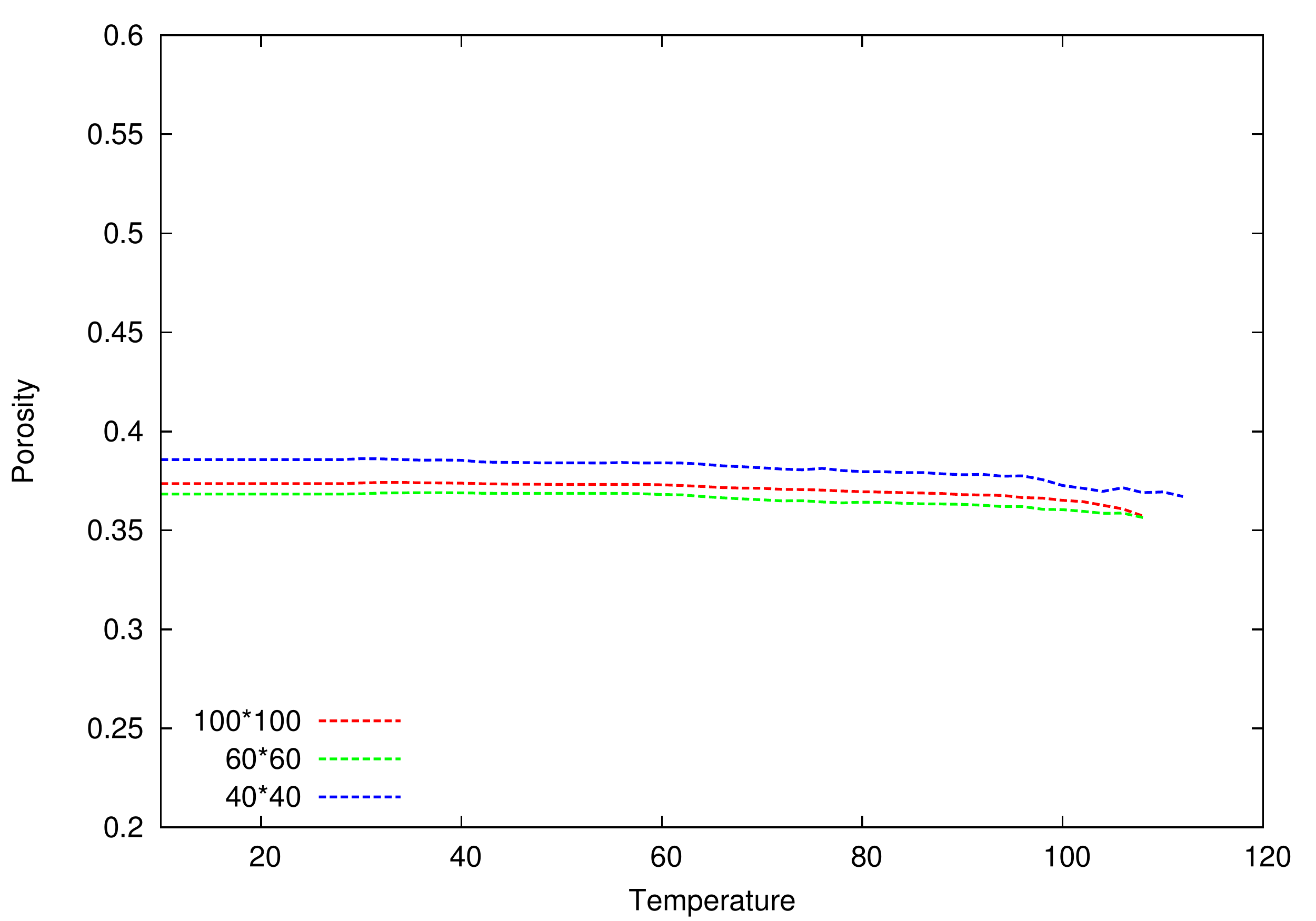}
\caption{Total surface area constituted by the pores (left panel) and porosity (right panel) as a function of temperature for different grid sizes 100$\times$100, 60$\times$60 and 40$\times$40. We define the porosity as a volume fraction (pore volume / total volume). Note that the total surface is scaled (divided by 4 and 2 for a 100$\times$100 and a 60$\times$60 grid respectively) to allow comparison.}
\label{vol}
\end{figure*}

The tags on the empty cells can directly be related to the volume of the pores.The cross section of the pores is estimated by adding the faces of tetrahedrons that are located between a pore and a water molecule. To do so, we select the empty cells which have water neighbours (they have a tag = 0) and check how many water molecules are around each empty cell. Each time a water molecule is next to an empty space in our grid, one surface area of a tetrahedron is added. Therefore, an empty space surrounded by four water molecules adds four faces of tetrahedron to the total surface. To calculate the total surface formed by the pores we therefore add all the surfaces at the interface between empty and filled cells. The total volume of the pores is calculated by adding the volume of the empty tetrahedrons. The total surface and the porosity (volume of the pores divided by total volume) are reported in figure ~\ref{vol} for different grid sizes. While the relative volume of the pores does not change drastically during warming up, the total surface decreases by a factor 3.5 between 10 and 120~K, influencing the increase in average pore size. This is consistent with a recent study from \cite{mitterdorfer2014} who show, by using small angle neutron scattering of ASW ice deposited at 77~K, that the specific surface area decreases from 90 to 100K. The pores present in our simulations reach these sizes at $\sim$100~K.The volume and surface of the pores calculated with different grid sizes, show similar behaviours as the temperature of the ice is increased. In this study, we consider the barrier for diffusion to be 40$\%$ of the binding energy. This barrier is uncertain, as mentioned before, and a change would imply that the decrease in surface area would start at lower temperatures for lower diffusion barrier, or higher temperatures for higher diffusion barriers. The change in this barrier would therefore change the temperature at which the re-arrangement of water molecules occurs, and pores coalesce.
 
\subsection{Comparison with experimental results.}
In the previous experimental studies (\citealt{isokoski2014}; \citealt{bossa2014}; \citealt{bossa2012}), several thousands of monolayers of water are deposited on a 2.5 x 2.5 cm slab. A clear finding is that the ice gets thinner for higher temperatures, and this cannot be concluded from the present simulations where the thermal decrease of thickness seems to stop for ices thicker than several tens of monolayers. This is likely an artifact that is caused by the nature of the small grid sizes that necessarily have to be used in the simulation. Whereas the large size of the experimental surface allows the molecules to diffuse through many different routes to reach a more stable configuration, this is not the case in the simulated ices, simply as the simulation boxes are too limited to allow a re-allocation of the ices on a large scale. Already for the deposition of 60 ML, as shown in figure ~\ref{nnpores}, the compaction is not well reproduced in the simulation; the ice clearly does not get thinner. Instead the water molecules, by re-arranging and becoming more bound, create pillar-like structures blocking the diffusion of species within the box. On a macroscopic scale, therefore, the present work should not be regarded as an attempt to reproduce the experimental results. For this larger boxes are needed that are currently out of reach in terms of computer time. However, at the level of individual molecules, our simulation boxes allow to follow the evolution of porosity in ASW ices. The results show that pores do remain within the ices upon warming up, which is consistent with the findings by \cite{isokoski2014}. The density of the ices in our simulations varies from 0.57 to 0.63 g cm$^{-3}$ during deposition. A value of 0.61 g cm$^{-3}$ has been found experimentally. The main conclusion is that thermal evolution of porous ices involves pore growth rather than eradication of pores, a process not considered so far and fully consistent with the experimental findings. The porosity of the ice can be obtained by comparing the average density of the ice $\rho$ to the density of the compact ice ($\rho_c$=0.94 g / cm$^{3}$, \citealt{dohnalek2003}) as $\phi$=1-$\frac{\rho}{\rho_c}$. The porosity of the ices derived experimentally is $\sim$ 32$\%$. With our simulations, we derive a porosity of $\sim$30-40$\%$ as shown in figure~\ref{vol}. Our simulations are therefore in good agreement with the density and porosity derived experimentally. \cite{raut2007} showed that for ice deposited at 30-40~K, the pores are very small (micropores) and are narrow (less than three molecular diameters wide). As the temperature increases, the decline of microporosity has been suggested to be due to pore clustering (until $\sim$90~K) and at higher temperatures (between 90 and 130~K), it has been attributed to pore collapse (\citealt{wu2010}). Our simulations show that microporosity decreases due to pore clustering. However, we could not simulate the pore collapse as found for high temperatures. In the present study, we do not model the effect of the deposition angle on the porosity of the ice as investigated by \cite{Kimmel2001} and \cite{dohnalek2003}. Initial density and porosity are indeed strongly dependent on the deposition angle as deposition angles above 40 degree create a filamentary structure within the ice. We do not observe a real pore connection network after deposition, mainly because there is no preferential orientation of the impinging water molecules. However, our model reproduces well the background deposition, since we obtain an initial density and an initial porosity close to the experimental results that started from low temperature background deposition.

\section{Discussion and Astronomical implications}
In this study we confirm that the amount of porosity of interstellar ices is not fully reflected by detections of OH dangling bounds. The surface area of the pores in ASW water ices decreases with increasing temperatures, which implies a decrease of the OH dangling bonds, while the total volume of the pores does not change significantly. We therefore conclude that a measurement of the OH dangling bonds does not give an unambiguous measure of the level of porosity of interstellar ices and we find that pores remain in ice upon warming up. This has already been discussed by \cite{isokoski2014} and \cite{raut2007} who showed experimentally that some porosity of the ices remains at high temperatures. Depending on the growth temperature, the residual porosity after annealing at 120~K is still around 20$\%$. 

\begin{figure}[h]
\includegraphics[scale=0.4,angle=-90]{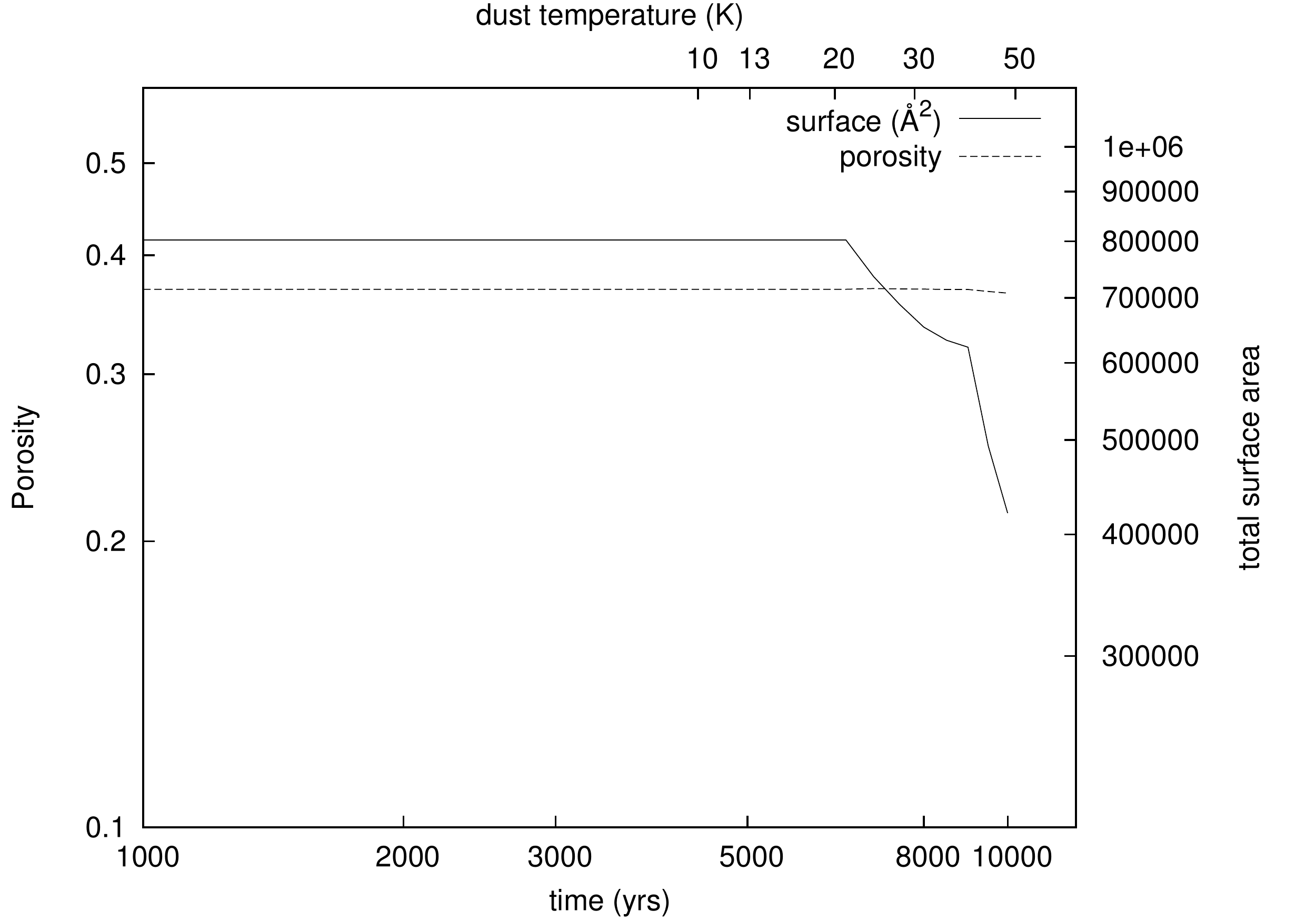}
\caption{ Evolution of the porosity in star forming environment. We assume dust temperature to increase from 10~K to 200~K in 5 10$^4$ years.}
\label{ism}
\end{figure}

In dense molecular clouds temperatures are around 10-14 K. In this case, the mechanism of re-structuring the ices would take time scales much longer than 10$^7$ years since it takes around t$\sim$$(\nu \exp{\frac{-0.4\times Ea}{T_{dust}}})^{-1}$$\sim$10$^{11}$ years for a water molecule (with one neighbour) to break and reform bonds at 14~K. Therefore the evolution of interstellar ices through diffusion should not be considerable in  molecular clouds. Once a star is born and heats up its environment, however, the dust grains become warmer and water molecules in the ice diffuse faster. We modelled the evolution of the porosity in star forming environments by using a heating schema similar to that used by \cite{garrod2006}. In this schema, the temperature of gas and dust, initially at 10~K, reaches 200~K after 5 $\times$ 10$^4$, 2 $\times$ 10$^5$ and 10$^6$ years, to mimic the thermal evolution of environments surrounding high-mass, intermediate, and low-mass star-formation, respectively. The increase of temperature follows time t as T=10+kt$^2$. In this study, we used the thermal evolution of a massive star, in 5$\times$10$^4$ years. In figure~\ref{ism}, we show the rough evolution of the volume and surface of the pores in an environment heated up by a newly formed star. We show that as temperature increases reaching 30~K, which corresponds to $\sim$8$\times$10$^3$ years, the porosity of the ice begins to change. The re-structuring occurs until temperatures of 50~K have been reached, which corresponds to a time scale of 10$^4$ years. In these time scales, with increasing temperature, pores connect to each other to form bigger entities. The chemical species trapped in these pores can encounter only when a certain temperature is reached. In this sense, ices could lock up many chemical species during long time scales and let them meet when conditions for reactions between these species are more probable. In this study, we used a heating schema from \cite{garrod2006} to mimic the temperature of the dust around a massive protostar. However, recent studies suggest that the heating schema could be much longer than the one considered here (\citealt{esplugues2014}). In this case, the decrease of surface area would occur at even lower temperatures than shown in  figure~\ref{ism}. Also, one should notice that the total surface area decreases by a factor of 2 in the ISM between 20 and 50 K while this decrease occurs between 28 and 90~K under laboratory conditions.These differences are due to the different time scales between the ISM and laboratory settings.

While we show that in star forming environments the pores present in interstellar ices could connect and form bigger pores, several previous studies have shown that pores may disappear during the molecular cloud phase because of radiation (\cite{palumbo2006} for cosmic rays). The corresponding timescale for ice compaction by a distribution of cosmic rays was extrapolated and demonstrated to be shorter (a few millions years) than the estimated lifetime of dense clouds, above a few 10$^7$ years. 

\begin{figure}[h]
\includegraphics[scale=0.3]{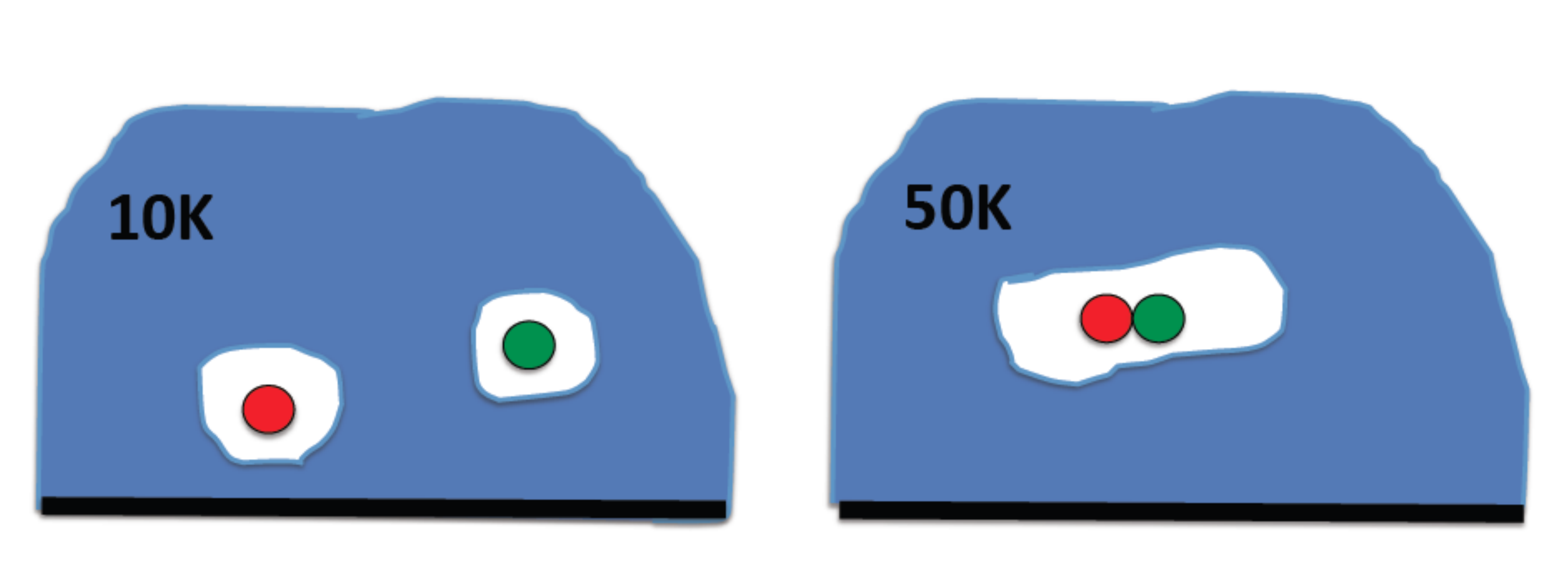}
\caption{ Sketch depicting the new mechanism we propose in this work to form species in ices. The species trapped in small pores at low temperatures (10~K) can react at higher temperatures when pores merge.}
\label{sketch}
\end{figure}

We show here that the evolution of interstellar ices as temperature increases is associated with the growth of pores instead of their eradication. Therefore if pores do survive the molecular cloud phase, they will evolve during warming up, grow and connect to form larger entities. This process could be of relevance for the formation of complex species as illustrated in figure~\ref{sketch}. This provides a new pathway for reactions taking place within the bulk of interstellar ices, and may boost the solid state formation of new molecules in space.




\begin{acknowledgements}
S. Cazaux is supported by the Netherlands Organization for Scientific Research (NWO). Part of this work was supported by NOVA, the Netherlands Research School for Astronomy, a Vici grant from the Netherlands Organisation for Scientific Research (NWO), and the European Community 7th Framework Programme (FP7/2007-2013) under grant agreement n.238258.  Support for J.-B. Bossa from the Marie Curie Intra-European Fellowship (FP7-PEOPLE- 2011-IEF-299258) is gratefully acknowledged. S.Cazaux would like to thank Dr. Thomas la Cour Jansen for very fruitful discussions. We would like to thank the referee for providing constructive comments which considerably improved our manuscript. 
\end{acknowledgements}

\clearpage

\clearpage

\end{document}